# Electronic Equivalent of a Mechanical Impact Oscillator

Volodymyr Denysenko[1,a], Marek Balcerzak[1], Artur Dabrowski[1]

## Affiliations

[1] Division of Dynamics, Lodz University of Technology, 90-924 Lodz, Poland

[a] Author to whom correspondence should be addressed: volodymyr.denysenko@dokt.p.lodz.pl

## Abstract

This paper presents a novel design of an electronic circuit that is equivalent to a mechanical discontinuous impact oscillator exhibiting hard impacts. The governing equations of the electronic circuit are derived to demonstrate its equivalence to the mechanical system. Numerical simulations of the electronic circuit are compared with those of the mechanical oscillator in both single and coupled configurations, showing a high degree of consistency between the two systems.

## Introduction

Piecewise-linear impact oscillator [1] is one of the simplest discontinuous mechanical systems. While the dynamics of a continuous linear oscillator are straightforward and predictable, the introduction of impacts enables the system to exhibit a wide range of complex behaviors, including period doubling and even chaotic solutions. Different modifications of this system can be applied, for example, as drive mechanisms [2] or energy harvesters [3]. Moreover, impact oscillators—particularly those exhibiting chaotic solutions—can be highly useful in studies of chaotic synchronization in discontinuous systems [4].

Although experiments with real mechanical oscillators provide the most direct means of investigating their dynamics, electronic analogues are often preferred [5] [6] [7]. This approach offers several notable advantages, especially in the context of synchronization studies. First, such research typically involves identical or nearly identical oscillators, which are much easier to construct using electronic circuits than mechanical systems [8]. Second, the implementation of coupling schemes—particularly asymmetric or master–slave configurations—is substantially more complex in mechanical setups, whereas in electronic systems, these can be realized with high precision and flexibility [9]. Another advantage is the ease with which system parameters can be adjusted in electronic circuits, using variable capacitors or resistors [10]. Finally, for impact oscillators, electronic implementations allow the realization of perfectly elastic collisions—a condition that is exceedingly difficult to reproduce in mechanical systems.

Different approaches to electronic impact oscillator circuit design can be found in the literature. The electric impact oscillator described in [11] models collisions with a spring rather than with a rigid wall. Another electronic impact oscillator was proposed in [12]; however, despite exhibiting similar dynamical behaviour, its dynamics cannot be described using the same equations as a mechanical impact oscillator. Therefore, it cannot be considered as an analogue to the mechanical one.

In this article, a novel electronic circuit is proposed that is equivalent to a mechanical oscillator with perfectly elastic hard collisions—i.e., one in which the sign of the oscillator's velocity is instantly reversed at impact [13] is proposed. The main advantage of the proposed design is that its dynamic behavior closely reproduces that of the mechanical system. To demonstrate this equivalence, the equations governing the electronic oscillator are derived. Furthermore, numerical simulations are presented for both the model of mechanical impact oscillator, described by system of differential equations, and electronic circuit, in both single and coupled configurations. The results confirm a high degree of consistency between the circuit and the original model of the mechanical system.

## *Mechanical impact oscillator*

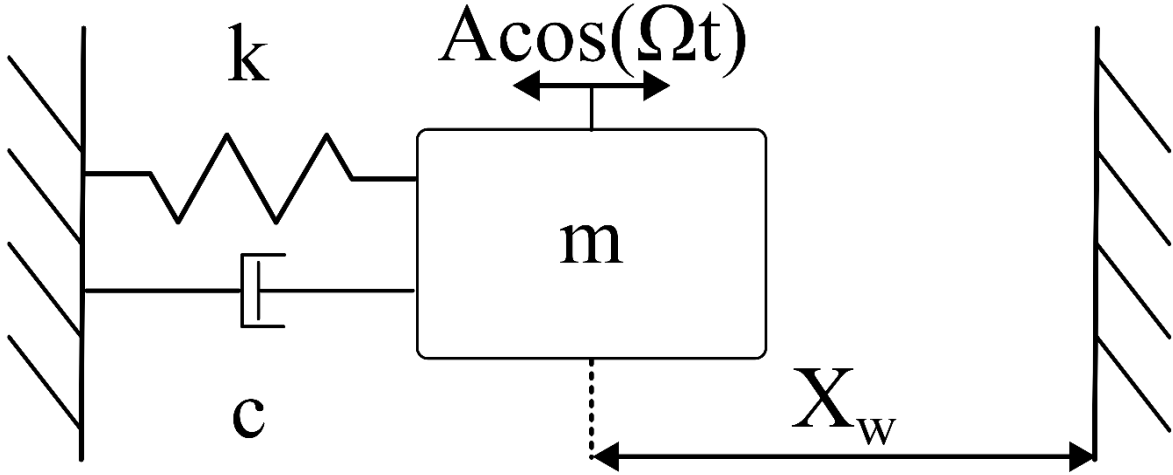

*Figure 1 Impact mechanical oscillator*

The mechanical impact oscillator (Figure 1), which is the typical mass-spring-damper system, that can impact the wall placed at the position $X_W$, is described by the following 2nd order differential equation:

$$m\ddot{X} + c\dot{X} + kX = A * cos(\Omega t) \qquad (1a)$$
$$\dot{X}(t_c^+) = -R * \dot{X}(t_c^-), X(t_c) = X_w \qquad (1b)$$

where $X$ is the displacement of the oscillator, m is its mass, $c$ is the damping coefficient, $k$ is the spring stiffness, $A$ and $\Omega$ are the amplitude and frequency of the external forcing, respectively. $R$ is the coefficient of restitution, and $t_c$ is the moment in time when the collision with the wall occurs. After reducing the Eq. (1) to the dimensionless form, the equations of motion are as follows:

$$x''(\tau) + 2\zeta x'(\tau) + x(\tau) = a * cos(\eta \tau) \qquad (2a)$$
$$x'(\tau_c^+) = -R * x'(\tau_c^-), x(\tau_c) = x_w \qquad (2b)$$

where $\zeta = \frac{c}{2\sqrt{mk}}$ is the damping ratio, $\omega = \sqrt{\frac{k}{m}}$ is the natural frequency of oscillator, $a = \frac{A}{m\omega^2}$, $\tau = \omega t$ is the dimensionless time, $\eta = \frac{\Omega}{\omega}$ is the dimensionless frequency of external excitation, $x = \frac{kX}{F}$, $x_w = \frac{kX_W}{F}$, and $\tau_c$ is the dimensionless time of collision.

## *Equivalent circuit*

The diagram of the designed circuit, which is equivalent to the mechanical impact oscillator described by Eq. (2), is shown in Figure 2. The circuit can be divided into two parts. The first part, comprising operational amplifiers (U1–U4) configured as inverting amplifiers or inverting integrators [14], corresponds to the equation of a smooth linear oscillator. Its dynamics are equivalent to those of an oscillator without impacts. The second part, which includes a comparator (U5, part number LT1394), a D-type flip-flop (A1), and two analog switches (U6, U7, part number ADG5233), handles impact detection and processing. The comparator (U5) detects impacts, the flip-flop (A1) stores the impact state, and the switches (U6, U7) change the velocity sign at the moment of impact.

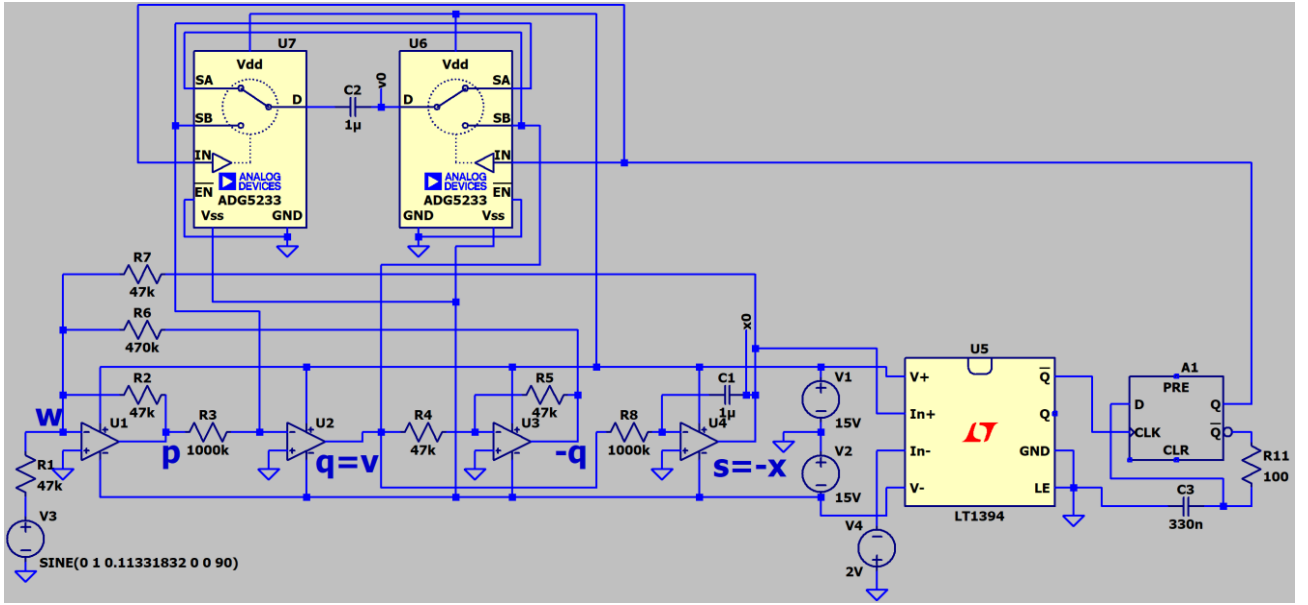

*Figure 2 The proposed electronic circuit, equivalent to Eq. (2)*

To demonstrate that the dynamics of the electronic circuit reproduce those described by Eq. (2), the circuit model must be derived in the form of a differential equation. To this end, the voltages at points **p**, **q**, and **s** must be considered. The voltage at point **p**, which is the output of operational amplifier U1 configured as a summing amplifier [14], is given by:

$$V_p = -R_2 \left( \frac{V_3}{R_1} - \frac{V_q}{R_6} + \frac{V_s}{R_7} \right) \tag{3}$$

where $V_3$ is the external source. Operational amplifiers U2 and U4 are configured as integrators; therefore, the rate of change of the voltage at points **q** and **s**, which are the outputs of U2 and U4 respectively, can be described as follows:

$$\dot{V}_q = -\frac{1}{R_3 C_2} V_p \tag{4a}$$

$$\dot{V}_s = -\frac{1}{R_8 C_1} V_q \tag{4b}$$

Substituting Eq. (3) into Eq. (4a) results in the following system of differential equations.

$$\dot{V}_s = -\frac{1}{R_8 C_1} V_q \tag{5a}$$

$$\dot{V}_q = -\frac{R_2}{R_3 C_2} \left( \frac{V_3}{R_1} - \frac{V_q}{R_6} + \frac{V_s}{R_7} \right) \tag{5b}$$

By differentiating Eq.(5a), the following expression is obtained.

$$-R_8 C_1 \ddot{V_s} = \dot{V_q} \tag{6}$$

Inserting Eq. (5a) and Eq. (6) into Eq. (5b) leads to the following differential equation of the 2$^{nd}$ order.

$$\ddot{V_s} = \frac{R_2}{R_8 C_1 R_3 C_2}\left(\frac{V_3}{R_1} + \frac{R_8 C_1 \dot{V_s}}{R_6} + \frac{V_s}{R_7}\right) \tag{7}$$

Assuming $V_s = -x, R_3 C_2 = R_8 C_1 = \omega^{-1} = 1, \frac{R_2}{2R_6} = \zeta, R_7 = R_2 = R_1, V_3 = V_A \cos(\Omega t)$, Eq.(7) can be rewritten in the following form.

$$\ddot{x} + 2\zeta\omega\dot{x} + \omega^2 x = V_A \omega^2 \cos(\Omega t) \tag{8}$$

Next, introducing non-dimensional time $\tau = \omega t$, Eq. (8) can be simplified as follows:

$$x''(\tau) + 2\zeta x'(\tau) + x(\tau) = V_A \cos(\eta\tau) \tag{9}$$

where $\eta = \frac{\Omega}{\omega}$ is non-dimensional frequency of the external excitation. As can be seen, Eq.(9) is identical to Eq.(2a), indicating that the proposed circuit is equivalent to a mechanical linear oscillator.

Next, the impact with the wall, described by the Eq.(2b), must be implemented in the circuit. To detect the equivalent of a collision with the wall, a voltage comparator U5 was applied to compare $V_s$, representing the negative position of the oscillator, with the reference voltage, applied to the inverting input $V_-$ of the U5 comparator, which corresponds to the position of the wall $x_w$. When the comparator detects a collision, it triggers a change in the state of a D-type flip-flop A1. This, in turn, alters the states of the analog switches U6 and U7, which are connected to capacitor C2. Since C2 is part of the integrator circuit based on operational amplifier U2, the switching action reverses the polarity of capacitor C2, thereby inverting the sign of the voltage $V_q$, which corresponds to the mechanical oscillator velocity. As a result, the proposed circuit effectively replicates the behavior of a mechanical impact oscillator with a coefficient of restitution $R = 1$.

## *Numerical verification*

To verify the equivalence between the proposed circuit design and the mechanical system described by Eqs. (2), the system trajectories obtained from the circuit were compared with those derived from numerical solutions of Eqs. (2). To ensure that the verification is both general and robust, system parameters were selected to represent cases exhibiting both chaotic and periodic behavior. The parameters of the system were picked from the article [15], where the largest Lyapunov exponent (LLE) was calculated over a range of parameters and a bifurcation diagram for the linear impact oscillator was presented.

The selected parameter values are: $\beta = 0.05, x_w = 2.0, R = 1.0$, while $\eta$ is considered as a bifurcation parameter. Three values of $\eta$ were chosen to represent distinct dynamical regimes: $\eta = 0.70$, which corresponds to the periodic solution without impact with the wall, $\eta = 0.712$, which corresponds to the chaotic solution, and $\eta = 0.74$, which corresponds to the period-3 solution.

To evaluate the suitability of the proposed circuit for investigating complete synchronization stability, we utilize a property of diagonally coupled systems described in [16]. A generalized system of two diagonally coupled arbitrary oscillators can be expressed as:

$$\dot{x} = f(x) + k_1 D(y - x) \tag{10a}$$

$$\dot{y} = f(y) + k_2 D(x - y) \tag{10b}$$

where $x \in \mathbb{R}^n$ and $y \in \mathbb{R}^n$ represent the state vectors of the first and second systems, respectively; $f: \mathbb{R}^n \to \mathbb{R}^n$ is the vector field describing the system dynamics; $D$ is a diagonal coupling matrix; and $k_1, k_2$ are the coupling coefficients. It has been shown that the systems tend toward complete synchronization if the sum of the coupling coefficients exceeds the largest Lyapunov exponent (LLE) of the uncoupled system, i.e., $k_1 + k_2 > \lambda_{LLE}$. Thus, the synchronization behavior—whether coherent or incoherent—depends on the LLE of the individual oscillator. This property can be exploited to validate the behavior of a system composed of two coupled electronic oscillators. To simplify the design of the two-oscillator circuit while preserving the described feature, it is proposed to use unidirectional coupling ($k_1 = 0, k_2 = k$). Corresponding system of dimensionless differential equations is as follows.

$$x_1'(\tau) = x_2(\tau) \qquad (11a)$$

$$x_2'(\tau) = -2\zeta x_2(\tau) - x_1(\tau) + a * \cos(\eta \tau) \qquad (11b)$$

$$x_2(\tau_c^+) = -Rx_2(\tau_c^-), x_1(\tau_c) = x_w \qquad (11c)$$

$$y_1'(\tau) = y_2(\tau) + k(x_1(\tau) - y_1(\tau)) \qquad (12a)$$

$$y_2'(\tau) = -2\zeta y_2(\tau) - y_1(\tau) + a * \cos(\eta \tau) + k(x_2(\tau) - y_2(\tau)) \qquad (12b)$$

$$y_2(\tau_c^+) = -Ry_2(\tau_c^-), y_1(\tau_c) = x_w \qquad (12c)$$

In order to simplify the design of the electronic circuit, equivalent to the Eqs. (12-13), this system can be reduced to the following pair of the second-order differential equations:

$$x''(\tau) + 2\zeta x'(\tau) + x(\tau) = a * \cos(\eta \tau) \qquad (13a)$$

$$x'(\tau_c^+) = -Rx'(\tau_c^-), x(\tau_c) = x_w \qquad (13b)$$

$$y''(\tau) + 2\zeta y'(\tau) + y(\tau) + k(1 + 2\zeta)(y'(\tau) - x'(\tau)) + 2k(y(\tau) - x(\tau)) = a * \cos(\eta \tau) \qquad (14a)$$

$$y'(\tau_c^+) = -Ry'(\tau_c^-), y(\tau_c) = x_w \qquad (14b)$$

Values of the system parameters in case of 2 coupled oscillators are the same as in case of the single oscillator with $\eta = 0.712$. The electronic circuit, equivalent to the system, described with the Eqs. (13-14) consists of 2 connected circuits and a coupling circuit, shown in Figure 3.

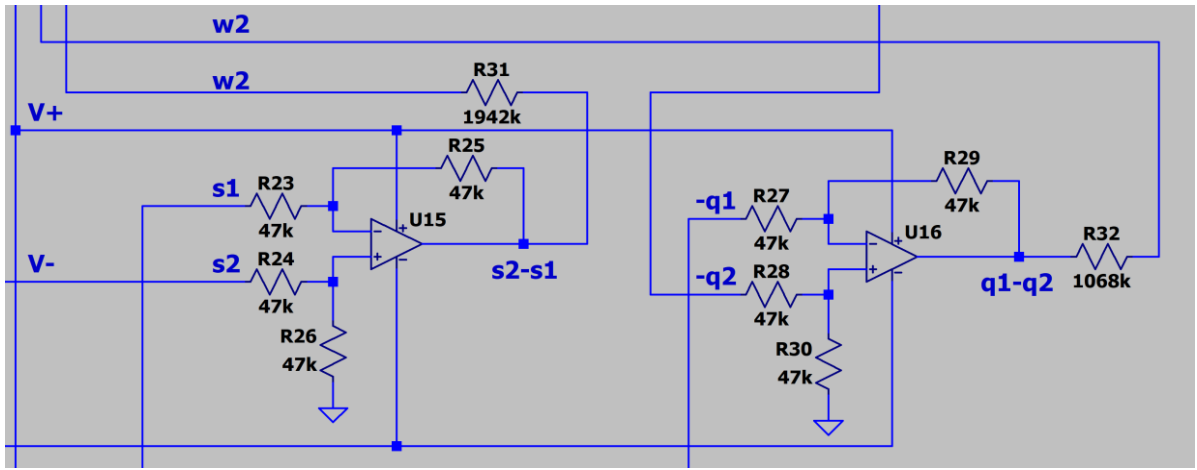

*Figure 3 Schematic of the electronic circuit implementing unidirectional diagonal coupling from Eq.(13-14)*

The coupling circuit is implemented using 2 differential amplifiers. Amplifier U15 corresponds to the position coupling term, while amplifier U16 corresponds to the velocity coupling term. The signal $V_{s1}$, representing the negative displacement of the first oscillator, is connected to the inverting input of U15, and $V_{s2}$, representing the negative displacement of the second oscillator, is connected to the non-inverting output. As a result, the output of the U15 is proportional to the $V_{s2} - V_{s1}$, which corresponds to $x - y$ in terms of oscillators displacement. Similarly, signals $V_{-q1}$ and $V_{-q2}$, representing the negative velocity of the first and second oscillator, respectively, are connected to the U16 in the same configuration as position counterparts. Thus, the output of the U16 is proportional to the $V_{q1} - V_{q2}$, corresponding to $x' - y'$ in terms of the oscillators velocities. To simplify the circuit design, it is assumed that $R_{23} = R_{24} = R_{25} = R_{26}$ and $R_{27} = R_{28} = R_{29} = R_{30}$, ensuring that U15 and U16 yield unamplified differences in position and velocity, respectively. The outputs of U15 and U16 are then connected to the summing amplifier of the second oscillator via resistors $R_{31}$ and $R_{32}$, respectively. By applying the same methodology used for the single oscillator, the dynamical model of the complete electronic circuit—comprising two oscillators and the coupling—can be derived. The terms corresponding to the dynamics of a single oscillator remain the same in the electronic circuit, as in the case of the circuit representing a single oscillator. The coupling terms, when implemented electronically, are represented as follows:

$$\frac{R_3}{R_{31}} = 2k; \quad \frac{R_3}{R_{32}} = k(1 + 2\zeta) \qquad (15)$$

The trajectories, both for the single and for the coupled oscillators, were obtained using analytical solutions. To detect collisions, trajectory of the oscillator was evaluated every $\Delta t = 10^{-3}$ of the external excitation period. If a condition $x(t + \Delta t) > x_w$ was met, the precise moment of impact was determined using the bisection method.

The proposed equivalent electronic circuits—comprising both the single oscillator and the coupled oscillator configurations—were simulated using LTspice software [17]. For the single oscillator, simulations were performed over a time span corresponding to 200 periods of external excitation. For the coupled oscillators, the simulation duration was extended to 500 periods to capture long-term synchronization behavior. Both circuits, fully prepared and configured for numerical simulations, are available in the repository [18].

## Results

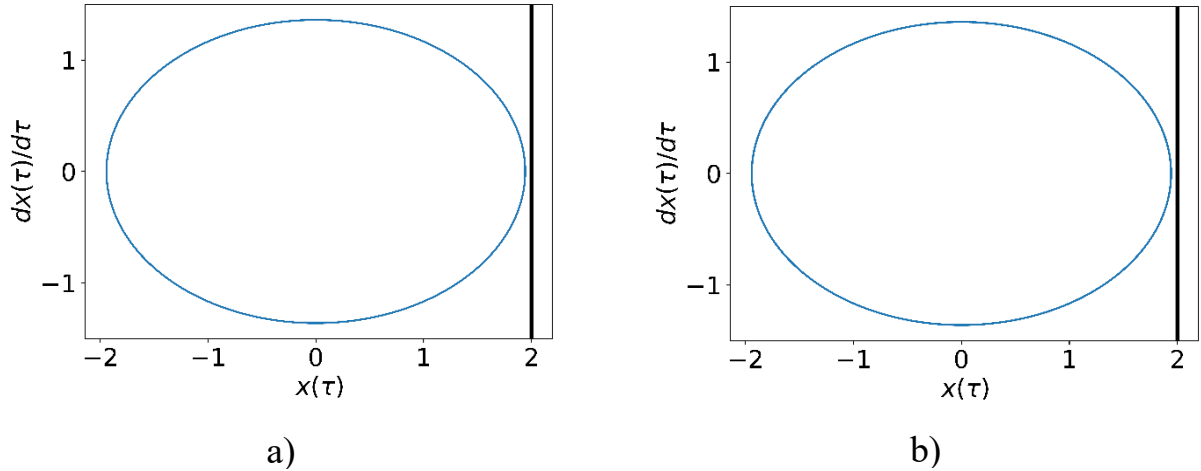

a)            b)

*Figure 4 Phase-space trajectory of the oscillator for $\eta = 0.7$, obtained using: a) numerical simulation, b) circuit simulation*

The phase-space trajectory of the single oscillator for $\eta = 0.7$, obtained numerically, is shown in Figure 4a, while the corresponding trajectory from the circuit simulation is presented in Figure 4b. It can be observed that, for this parameter value, the oscillator does not attain sufficient amplitude to reach the wall after the transient phase. The trajectories obtained by both methods are in excellent agreement, confirming the accuracy of the proposed circuit model.

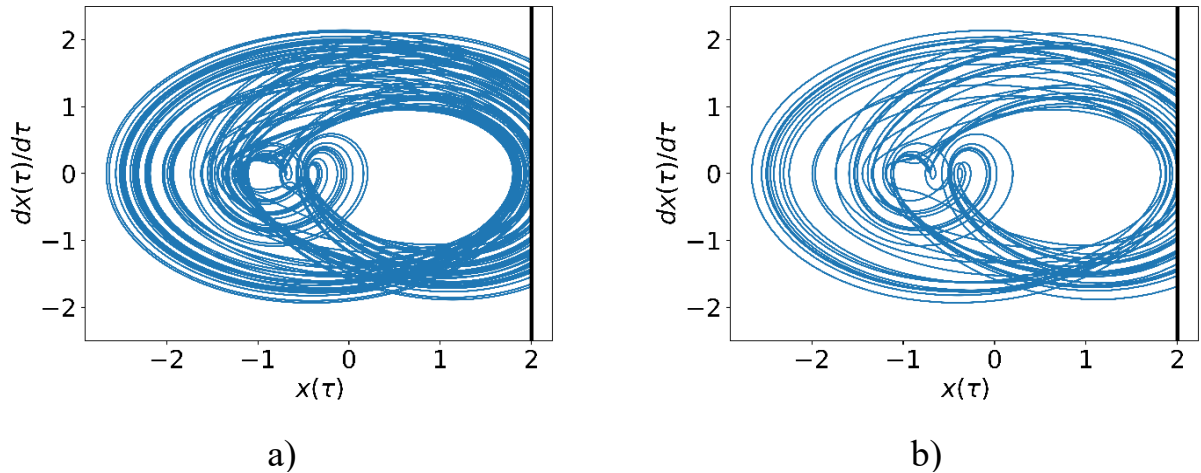

a)            b)

*Figure 5 Phase-space trajectory of the oscillator for $\eta = 0.712$, obtained using: a) numerical simulation, b) circuit simulation*

Figures 5a ansd 5b present the phase-space trajectories of the impact oscillator for $\eta = 0.712$, obtained through numerical simulation and circuit simulation, respectively. Both methods yield a chaotic attractor, as expected for this parameter value. However, noticeable differences between the attractors are evident. These discrepancies are likely attributable to numerical inaccuracies inherent in the circuit simulation, which are amplified by the extreme sensitivity of chaotic systems to initial conditions and small perturbations.

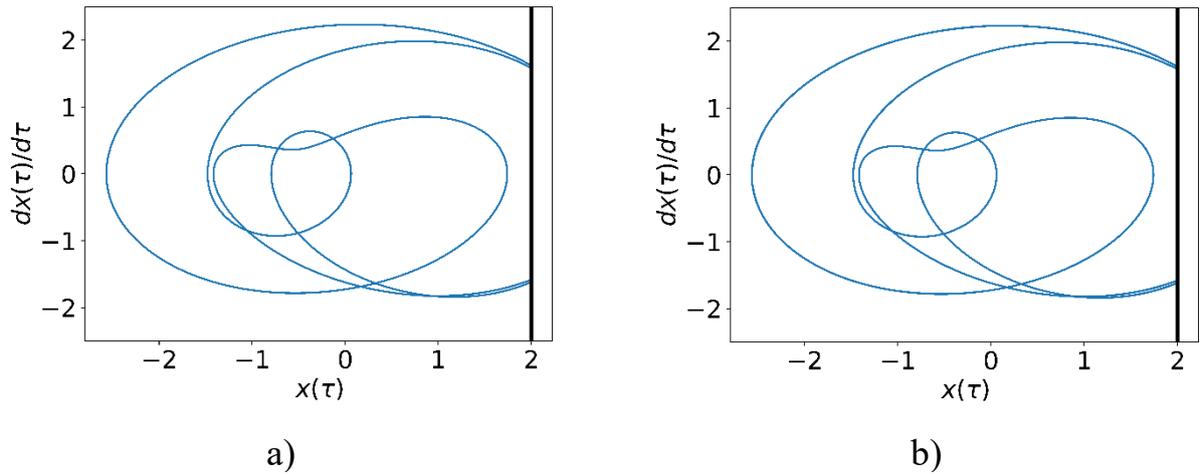

Figure 6 Phase-space trajectory of the oscillator for $\eta = 0.74$, obtained using: a) numerical simulation, b) circuit simulation

Figure 6a displays the phase-space trajectory of the single oscillator for $\eta = 0.74$, obtained through numerical simulation, while Figure 6b presents the corresponding trajectory from the circuit simulation. In both cases, the oscillator exhibits a period-3 solution. The resulting trajectories are virtually identical, further validating the accuracy of the proposed electronic circuit in replicating the dynamics of the mechanical impact oscillator.

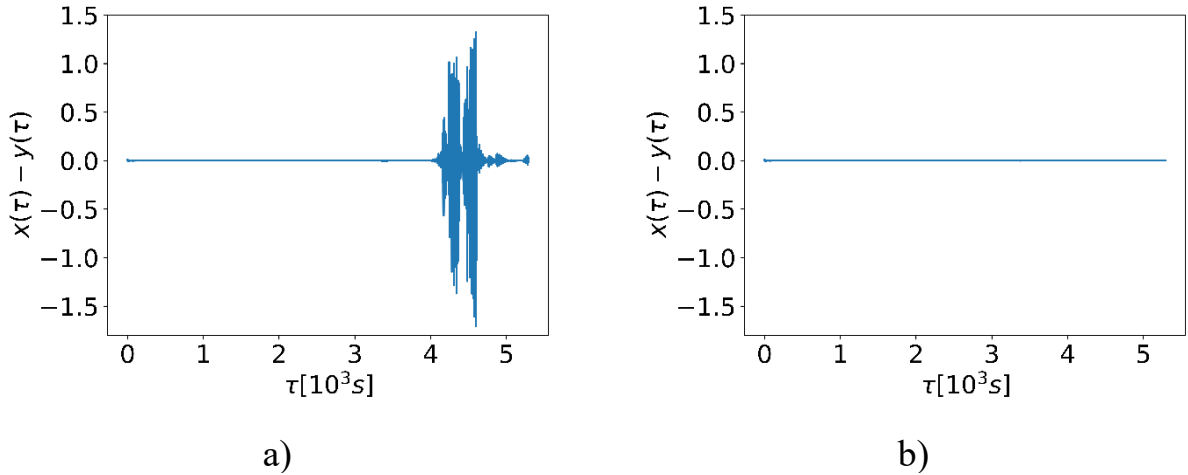

Figure 7 Time series of the position differences between two unidirectionally coupled oscillators for coupling coefficients: a) $k = 0.031$, b) $= 0.032$

Figures 7a and 7b illustrate the time series of the position differences between two unidirectionally coupled oscillators for coupling coefficients $k = 0.031$ and $k = 0.032$, respectively. As shown, the oscillators fail to synchronize in the first case, while stable synchronization is achieved in the second. This indicates that the largest Lyapunov exponent (LLE) of the single oscillator lies between these two values, i.e., $0.031 < \lambda_{LLE} < 0.032$. The LLE value calculated using the numerical method described in [19] is approximately equal to $\lambda_{LLE} \approx 0.03$. Considering that the method used to estimate the LLE from the electronic oscillator is less precise than analytical alternatives, the observed result is within an acceptable margin of error. Another factor that might have contributed to the aforementioned error is the inaccuracy of the LTspice solver.

## Discussion

Based on the comparison between the trajectories obtained from the circuit simulation of the single oscillator and those derived from numerical simulations, it can be concluded that the trajectories are nearly identical in cases where the system exhibits periodic behavior. Discrepancies appear only in the chaotic regime, which can be attributed to both the inherent sensitivity of chaotic systems and numerical inaccuracies introduced by the electronic circuit simulator. Nevertheless, the qualitative behavior of the equivalent circuit remains consistent with that observed in the numerical simulations.

Furthermore, the LLE of the equivalent circuit was estimated using the synchronization-based method. Although the obtained value differs slightly from the LLE calculated via numerical and analytical methods, the deviation is acceptable given that the more precise analytical estimation was applied to the numerical model of the impact oscillator, while the synchronization method was used for the electronic implementation.

Overall, the results presented for both the single oscillator and the system of two diagonally coupled oscillators demonstrate that the proposed circuit accurately replicates the dynamics of the mechanical piecewise-linear impact oscillator. Therefore, it can be effectively utilized for experimental investigation of such systems and their complex behaviors.

## Acknowledgements

This paper has been completed while the first author was the Doctoral Candidate in the Interdisciplinary Doctoral School at the Lodz University of Technology, Poland.

## Author Declarations

## Conflict of interest

The authors have no conflicts to disclose

## Funding

V.D has been supported by the National Science Centre, Poland, PRELUDIUM Program (Project No. 2023/49/N/ST8/02436)

## Author contributions

**V.D.:** Investigation, Methodology, Writing – original draft, Visualization, Writing – review & editing; **M.B.:** Conceptualization, Investigation, Supervision, Writing – review & editing; **A.D.:** Funding Acquisition; Supervision

## Data availability

All circuit files prepared and configured for numerical simulations, using LTspice, along with the script developed in this study, are available in the research data linked to this paper *[18]*.